\documentstyle[12pt,twoside]{article}

\begin{document}
{
\title{\Large\bf ZERO TEMPERATURE BLACK HOLES AND\\
THE FAILURE OF SEMI-CLASSICAL ANALYSIS}

\author{\Large F.G. Alvarenga, A.B. Batista,
J. C. Fabris and G.T. Marques\\
Departamento de F\'{\i}sica, Universidade Federal do Esp\'{\i}rito Santo, \\
CEP29060-900, Vit\'oria, Esp\'{\i}rito Santo, Brazil\\
} 
\maketitle

\begin{abstract}
The extreme Reissner-Nordstr\"om black holes have zero surface gravity. However, a semi-classical
analysis seems to be ill-definite for these objects and apparently no notion of temperature
exists for them. It is argued here 
that these properties are shared for all kind of
black holes whose surface gravity is zero. Two examples are worked out explicitely:
the scalar-tensor cold black holes and extreme black holes resulting from a gravity system
coupled to a
generalized Maxwell field in higher dimensions.
The reasons for this anomolous behaviour are discussed as well as its
thermodynamics implications.
\end{abstract}

The temperature of a black hole may be obtained in different ways. It is generally
stated that this temperature is proportional to the black hole surface gravity.
When the surface gravity is not zero,
this statement is consistent with the temperature obtained through other methods, like the euclideanization of
the metric or the computation of the Bogolubov's coefficients \cite{hawking1,birrel,spindel,mitra}.
However, these considerations seem to be more delicate for those black holes
whose surface gravity is zero. In the case of the Reissner-Nordstr\"om  solution, the extreme black holes
have zero surface gravity suggesting that they have zero temperature. However, the computation
of the temperature through the euclideanization of the metric indicates an arbitrary value \cite{mitra,mitra2,hawking}; moreover, the computation of the Bogolubov's coefficients seems to be
ill-definite in the sense that the normalization conditions are not satisfied indicating
a break down of semi-classical analysis \cite{glauber}. Hence, it seems that no notion
of temperature can be assined to the RN extreme black holes.
\par
In this work, we will employ the term "extreme black holes" to denote those black holes
which are, in principle, the limit case of a more general class of black holes,
like in the Reissner-Nordstr\"om solution. There are black holes which have zero surface
gravity but do not match in this classification, like the so-called scalar-tensor
black holes \cite{kirill1,kirill3}: in this case, all possible black hole solutions have zero
surface gravity.
\par
Zero surface gravity are indeed very particular objects. In the case of the Reissner-Nordstr\"om
solutions, they are the limiting case as $M \rightarrow Q$. However, it has been argued in
\cite{hawking} that the RN extreme black holes must be seen as a qualitatively different objects
with respect to the non-extreme one, and not simply as the their limiting case:
an extreme black hole could not result from the evaporation of a non-extreme one, neither
they could not originate a non-extreme black hole by absorbing energy. In this sense, the
thermodynamics of the extreme black holes has many special features. For example, it is argued
that extreme black holes has zero entropy, even if its event horizon area is not zero \cite{hawking, zaslavskii}. Hence
the law $S = {\cal A}_{BH}/4$ would be not valid for these holes. However, computation of
the entropy by counting the number of state in the context of string theory leads to the
opposite result \cite{horowitz}.
\par
In a previous work \cite{glauber}, it has been shown that the Bogolubov's
coefficients is not well definite for the extreme black hole, and no notion of
temperature can be obtained in this way: the Bogolubov's coefficients
obtained using a thin shell collapsing model do not obey the required normalization
conditions. This indicates a break down of the semi-classical approximation for the extreme
black holes. It is not clear if the notion of temperature for these holes can be recovered
if an exact analysis is employed instead (if such exact analysis is somehow feasable). One
feature of the extreme black holes that indicates that the notion of temperature
is not possible comes from the euclideanization of the metric: as it has been said before,
using this procedure, the
temperature of the extreme black hole is arbitrary (in any case, different from zero), what
could be re-stated by saying that it is not possible to assign to them any temperature at
all. This results comes from the fact that in the space-time of this hole, any point
outside the event horizon is at an infinite distance from it and due to this
the peridiocity of the euclidean time is arbitrary. This property is not shared
by the non-extreme black holes.
\par
The failure to perform a semi-classical analysis for the RN extreme black
hole, as pointed out in reference \cite{glauber}, comes from the relation between the
ingoing modes $u$ and outgoing modes $v$, when a collapsing model is worked out. While
in the non-extreme case, this relation is logarithmic, in the extreme case is a power law.
As a consequence, the Bogolubov's coefficients are very distinct. In particular, the
Bogolubov's coefficient $\beta_{\omega\omega'}$, which is connected with the particle
creation, computed for the non-extreme case goes to zero when the surface gravity goes
zero; but, imposing the extreme condition
from the begining, this coefficient is not zero, and predicts an infinite number of
particles created \cite{liberati1,liberati2}. More seriously, the normalization conditions, which indicate
the consistency of the semi-classical analysis, are not satisfied. Hence, the semi-classical
analysis breaks down in the extreme case, while is perfectly valid for the non-extreme case.
The fact that the limit of the non-extreme case when the surface gravity goes to zero
does not coincide with the calculations performed by imposing the extreme condition
from the begining support the statement of \cite{hawking}.
\par
The aim of the present paper is to show that for any black hole with zero surface gravity,
the semi-classical analysis must break down, and in principle there is no notion of temperature
for them. In particular, for such black holes the spatial distance of any point to the event horizon
is infinite. Hence, following the reasoning of reference \cite{hawking}, the temperature is
arbitrary. In other words, there is no notion of temperature for them. Moreover, the relation
between the ingoing and outgoing modes for any black hole with zero
surface gravity is not logarithmic (for which case, a semi-classical
analysis predicts a well-definite temperature), following instead a power law.
\par
Let us first recall the computation of the Bogolubov's coefficient for the non-extreme
and for the extreme RN black holes. In what follows, we have in mind
a simplified version of the collapsing model, considering a thin charged shell \cite{ford}. Initially,
the space-time is minkowskian; later, after the formation of the event horizon, the
space-time is asymptotically flat. The procedure consists in computing the Bogolubov's
coefficients $\alpha_{\omega\omega'}$ and $\beta_{\omega\omega'}$ between the initial
and final modes of a quantum field, which we consider as a massless scalar field for
simplicity. To details of this computation, we address the reader to reference \cite{glauber}.
\par
The metric for the Reissner-Nordstr\"om solution is,
\begin{equation}
\label{m1}
ds^2 = \biggr(1 - \frac{r_+}{r}\biggl)\biggr(1 - \frac{r_-}{r}\biggl)dt^2 - \biggr\{\biggr(1 - \frac{r_+}{r}\biggl)\biggr(1 - \frac{r_-}{r}\biggl)\biggl\}^{-1}dr^2 - r^2d\Omega^2
\quad ,
\end{equation}
where $r_\pm = M \pm \sqrt{M^2 - Q^2}$.
For the non-extreme case, the relation between the {\it in} and {\it out} modes is
given by \cite{hawking1,birrel,ford}
\begin{equation}
\label{rela}
u = - \kappa^{-1}\ln\biggr[\frac{v_0 - v}{C}\biggl] \quad ,
\end{equation}
where $\kappa$ is the surface gravity, $v_0$ is the last mode scattered before the
formation of the event horizon and $C$ is a constant which depends on some
parameters characterizing the collapse of the shell.
The Bogolubov's coefficient are obtained through the inner product between the {\it in}
and {\it out} modes:
\begin{equation}
\label{coef}
\alpha_{\omega\omega'} = (g_\omega,f_{\omega'}) \quad , \quad \beta_{\omega\omega'} = - (g_\omega,f^*_{\omega'}) \quad ,
\end{equation}
where the inner product is defined as
\begin{equation}
(\phi_1,\phi_2) = - i\int d\Sigma^\mu\biggr(\phi_1\partial_\mu\phi^*_2 - \partial_\mu\phi_1\phi_2^*\biggl) = -i\int d\Sigma^\mu\phi_1\stackrel{\leftrightarrow}{\partial}_\mu\phi_2^* \quad .
\end{equation}
Using the relation (\ref{rela}), the Bogolubov's coefficients are
\begin{eqnarray}
\label{bc1}
\alpha_{\omega\omega'} &=& \frac{2\sigma\omega}{2\pi\sqrt{\omega\omega'}}\Gamma(2i\sigma\omega)
e^{\sigma\pi\omega}\exp[i(\omega'v_0 - 2\sigma\omega\ln C - 2\sigma\omega\ln\omega')] \quad ,\\
\label{bc2}
\beta_{\omega\omega'} &=& - \frac{2\sigma\omega}{2\pi\sqrt{\omega\omega'}}\Gamma(2i\sigma\omega)
e^{-\sigma\pi\omega}\exp[-i(\omega'v_0 + 2\sigma\omega\ln C + 2\sigma\omega\ln\omega')] \quad .
\end{eqnarray}
In these expressions $\sigma = \kappa^{-1}$. Notice that when $\kappa \rightarrow 0$, the
$\beta_{\omega\omega'}$'s coefficient goes to zero, what is compatible with the
notion of zero temperature. From these coefficients we can obtain the temperature for a non-extreme
black hole:
\begin{equation}
T = \frac{1}{8\pi M}\biggr(1 - \frac{16\pi^2Q^4}{A^2}\biggl) \quad ,
\end{equation}
where $A = 4\pi r_+^2$ is the area of the event horizon. It can be verified that
when $Q \rightarrow M$, $T \rightarrow 0$.
\par
Let us impose now the extreme condition from the begining. Remark that
the extreme condition $M = Q$ in (\ref{m1}) implies that the
event horizon is degenerated. In this case, the relation
between the {\it in} and {\it out} modes is \cite{liberati1,liberati2,glauber}
\begin{equation}
\label{relb}
u = \frac{C}{v_0 - v} \quad .
\end{equation}
The Bogolubov's coefficient read now \cite{glauber},
\begin{eqnarray}
\label{bc3}
\alpha_{\omega\omega'} &=& \pm \frac{e^{i\omega'v_0}}{\pi}\sqrt{\frac{D}{\omega\omega'}}K_1(\pm 2i\sqrt{D}) = \mp \frac{e^{i\omega'v_0}}{\pi}\sqrt{\frac{D}{\omega\omega'}}H_1^{(1)}(\mp 2\sqrt{D})\quad , \\
\label{bc4}
\beta_{\omega\omega'} &=& \pm i\frac{e^{-i\omega'v_0}}{\pi}\sqrt{\frac{D}{\omega\omega'}}K_1(\pm 2\sqrt{D}) \quad ,
\end{eqnarray}
where $D = C\omega'\omega$. Note that the expressions (\ref{bc3},\ref{bc4}) are not the limit
of (\ref{bc1},\ref{bc2}) when the surface gravity goes to zero. This is consequence of
the fact that (\ref{relb}) is not the limiting case of (\ref{rela}) when $\kappa \rightarrow 0$.
\par
In order these computations to be consistent, the normalization condition
\begin{equation}
\label{c3}
\int_0^\infty\int_0^\infty\biggr\{\alpha_{\omega\omega'}\alpha^*_{\omega''\omega'} -
\beta_{\omega\omega'}\beta^*_{\omega''\omega'}\biggl\}d\omega'd\omega'' = 1 \quad ,
\end{equation}
must be obeyed. This occurs for the non-extreme case, while for the extreme case such
condition is violated. This implies that for the extreme black hole, the semi-classical
analysis breaks down. The question is: can the temperature of an extreme black hole be
obtained if an exact analysis is performed? First of all, it is not clear that technically
such an exact analysis can be performed. Second, the fact that the temperature for the extreme case obtaining by using
the euclideanization of the metric is arbitrary suggests that even if an exact analysis would be
possible, the anomaly pointed out above would remain. From these considerations, it seems
that a determinant feature for the failure of the semi-classical analysis and for obtaining
the notion of temperature for the extreme black hole is related to the infinite
spatial distance of any point to the event horizon.
\par
The spatial distance of any point to the event horizon is infinite always the surface gravity is
zero. To prove this assertion, let us consider a two dimensional static spherically symmetric model, where the metric takes the
form
\begin{equation}
ds^2 = g_{00}(r)dt^2 - \frac{dr^2}{g_{00}(r)} \quad .
\end{equation}
It is always possible to find a radial coordinate $r$ where the metric takes
this form, at least near the horizon. Now, the event horizon is characterized by $g_{00}(r_h) = 0$. Moreover,
a zero surface gravity means $g'_{00}(r_h) = 0$ \cite{wald}, where lines mean differentiation
with respect to $r$. The spatial distance to the event horizon
is given by
\begin{equation}
l = \int_{r_h}^r\frac{d\rho}{\sqrt{g_{00}(\rho)}} \quad .
\end{equation}
Near the horizon, we can perform a Taylor expansion, defining $x = r - r_h$. Considering zero surface gravity, we
find near the horizon
\begin{equation}
l \sim \int_{r_h}\frac{dx}{\sqrt{g''_{00}(r_h)x^2}} \sim \ln x|_{r_h} \rightarrow \infty \quad .
\end{equation}
Note that when the surface gravity is not zero ($g'_{00}\neq 0$), the integral converges.
\par
Now, in order to support the statement that the notion of
temperature is ill-definite for any black hole with zero surface gravity, two other particular cases will be analysed: the scalar-tensor cold black holes \cite{kirill1} and a class
of black holes obtained when gravity is coupled to a conformal field in higher dimensions \cite{kirill2}.
For the scalar-tensor black holes, let us consider those coming from the Brans-Dicke theory.
There are black holes solutions, in this case, only for the anomolous theory, that is, when
the Brans-Dicke parameter is such that $\omega < - 3/2$.
They admit, besides the mass, a scalar charge. They are called "cold black holes" because their
temperature, following from the evaluation of the surface gravity, is zero
irrespectivelly of the value of the scalar charge. What happens if we try to extract
the temperature from a semi-classical analysis? This mounts to evaluate the relation between the
{\it in} and {\it out} modes, obtaining the Bogolubov's coefficients.
\par
The metric describing the cold black holes is \cite{kirill1}
\begin{equation}
\label{m2}
ds^2 = \rho^{n+2}dt^2 - \frac{4k^2(m - n)^2}{(1 - P)^4}\rho^{-(n+2)}d\rho^2 - \frac{4k^2}{(1 - P)^2}\rho^{-m}d\Omega^2 \quad ,
\end{equation}
where
\begin{equation}
P \equiv 1 - \frac{2k}{r} \equiv \rho^{m - n} \quad .
\end{equation}
The event horizon appears at $\rho = 0$.
The parameters $m$ and $n$ are connected with the mass and the scalar charges. They must
be positive integers and must also satisfy the relation $m - n \geq 2$. These conditions
appear in the extension of metric, revealing its regularity.
The relation between the $u$ and $v$ modes is obtained using the
same procedure outlined before. In analysing the collapsing model,
there is in principle one difficulty arising from the fact that the scalar-tensor theory does
not obey the Birkhoff's theorem. But, since the only delicate point due to this could come
from the first crossing, when the metric is almost flat, such violation of the Birkhoff's theorem does not spoil the main steps of the calculation. Hence, we can proceed as it is
done in reference \cite{glauber}.
\par
Performing all the calculations, which implies to consider the metric (\ref{m2}) near the
horizon, we obtain the relation
\begin{equation}
u = \frac{C}{(v_0 - v)^{(1 + n)/(m - n)}} \quad .
\end{equation}
When $(1 + n)/(m - n) = 1$, that is $m = 2n + 1$, the relation $u = u(v)$ is the same as in the
RN extreme case. Indeed, for this case, when $n$ is even, the space-time diagram is
the same as in the Reissner-Nodstr\"om extreme case. When $n$ is odd, the structure of the
space-time is more involved, and it can admit even closed geodesics. Anyway, the problems presented before are the same as in the
RN extreme case, and no notion of temperature can be obtained: the semi-classical analysis
is simply not valid. We can only repeat what was done in reference \cite{glauber}.
\par
For the other cases where $(1 + n)/(m - n) \neq 1$, the Bogolubov's coefficients depends
on the particular values this ratio may take. In general, they are expressed in terms
of Meijer's functions. But, the coefficient $\beta_{\omega\omega'}$ is never zero, and
the same problems existing in the RN extreme case appear: the semi-classical analysis breaks
down. Note that the metric (\ref{m2}) leads to an infinite spatial distance of any point to the
event horizon, as we could expect since the surface gravity is zero. In reference
\cite{zaslavskii}, the evaluation of the entropy for these
scalar-tensor black holes was carried out: the author of this
reference find that these black holes have zero entropy, implying again a violation
of the usual entropy law for black holes, since they
have infinite horizon area.
\par
Another case, is the black holes in a muldimensional system where gravity is coupled to
a conformal maxwellian field. This is possible only if the dimension of the space-time is even.
The lagrangian for this system is \cite{fabris}
\begin{equation}
L = \sqrt{g}\biggr[R + (-1)^{d-1}\,F_{A_1....A_{\frac{d}{2}}}F^{A_1....A_{\frac{d}{2}}}\biggl]
\quad ,
\end{equation}
where $d$ here denotes the dimension of the space-time.
The static spherically symmetric solution in $d$ dimension with magnetic and electric
charges reads \cite{kirill2}
\begin{equation}
\label{m3}
ds^2_d = \frac{(r - r_+)(r - r_m)}{(r + r_e - r_m)^2}dt^2 - \frac{r^2\,dr^2}{(r - r_+)(r - r_m)}
-r^2d\Sigma^2 + \sum_{i=1}^s\,ds_i^2 \quad ,
\end{equation}
where 
\begin{equation}
r_+ = 2k + r_m \geq r_m \quad , \quad r_{e,m} = \sqrt{k^2 + 2q^2_{e,m}} - k \quad ,
\end{equation}
$k$ being again a parameter connected to the mass, $q_{e,m}$ denote the electric and
magnetic charges and $s$ is the number of Ricci-flat internal spaces. When $r = r_+$, an
event horizon appears. There is an extreme black hole  when $r_+ = r_m$: as in the
RN extreme case the event horizon is also degenerated here.
\par
Following the same procedure as before, we establish the relation between $u$ and $v$ by
employing a two-dimensional model of a collapsing thin charged shell.
For the non-extreme case ($r_+ > r_m$)
the relation reads,
\begin{equation}
u = - 2\frac{r_+(r_+ + r_e - r_m)}{r_+ - r_m}\ln\biggr[\frac{v_0 - v}{C}\biggl] =
- \kappa^{-1}\ln\biggr[\frac{v_0 - v}{C}\biggl] \quad .
\end{equation}
The evaluation of the Bogolubov's coefficients and, consequently, the temperature of these
holes follows the same steps as in the RN non-extreme case. The temperature reads
\begin{equation}
T = \frac{1}{4\pi}\frac{r_+ - r_m}{r_+(r_+ + r_e - r_m)} \quad .
\end{equation}
\par
However, for the extreme case $r_+ = r_m$, we now have
\begin{equation}
u = \frac{C}{v_0 - v} \quad ,
\end{equation}
and everything behaves as in the RN extreme case, with the same pathologies and inadequacy
of the semi-classical analysis. Note that for the extreme case treated in this
example, any spatial distance
to the event horizon is also infinite, in contrast with what happens with the
corresponding non-extreme case.
\par
In this work, we have discussed the notion of temperature and the using of semi-classical analysis for black holes whose
surface gravity is zero. In principle, these objects should exhibit a zero temperature.
But, the semi-classical analysis fails to apply to them. This problem has already been
treated in reference \cite{glauber} for RN extreme black holes. Our aim here
was to call attention that the anomaly present in the last case is much broader: any zero
surface gravity black hole seems to lead to the break down of the semi-classical analysis. The question that
arises in connection with this result is the precise reason for this. A common feature of all black holes with zero surface gravity is
the fact that the spatial distance of any point external to the horizon is infinite.
This implies that zero surface black holes have arbitrary temperature when the
euclidean metric method is used. In this approach,
the euclideanized metric is not regular near the horizon; so, any periodicity can
be assigned to the euclidean time. Mathematically, this feature is due
to the behaviour of the tortoise coordinate $r^*$ near the horizon: the dominant
term is a inverse power law instead of a logarithmic term \cite{mitra2}. Combined with the results reported here, all this seems
to indicate simply that such holes admit no notion of temperature. If we take into
account the results of reference \cite{hawking}, which reveals the failure of the usual
law of entropy for zero surface gravity black holes, it seems that all thermodynamics
of such holes is ill-definite. Note that this fact is somehow consistent with the
restrictions coming from the third law of thermodynamics \cite{israel}.
\newline 
\vspace{0.5cm}
\newline
{\bf Acknowledgements:} This work has been
partially supported by CNPq (Brazil) and CAPES (Brazil).

\end{document}